\documentclass[preprint2]{aastex}

\usepackage{verbatim}
\usepackage{natbib}
\usepackage{threeparttable}

\newcommand*{\lam}{\lambda}
\newcommand*{\Lam}{\Lambda}

\newcommand{\der}[2]{\frac{d #1}{d #2}}

\begin{document}

\title{Clearing the Gas from Globular Clusters \& Dwarf Spheroidals with Classical Novae } 
\author{Kevin Moore\altaffilmark{1} and Lars Bildsten\altaffilmark{1,2}}

\altaffiltext{1}{Department of Physics, Broida Hall, University of California, Santa Barbara, CA 93106, USA}
\altaffiltext{2}{Kavli Institute for Theoretical Physics, Kohn Hall, University of California, Santa Barbara, CA 93106, USA}

\begin{abstract}
Observations of the intra-cluster medium (ICM) in galactic globular clusters (GCs) show a systematic deficiency in ICM mass as compared to that expected from accumulation of stellar winds in the time available between galactic plane crossings. In this paper, we reexamine the original hypothesis of Scott and Durisen that hydrogen-rich explosions on accreting white dwarfs, classical novae (CNe), will sweep out the ICM from the cluster more frequently than galactic plane crossings. 
From the CNe rate and stellar mass-loss rate, this clearing mechanism predicts that $\approx 0.03\ M_\odot$ should be present in $\le 10^5M_\odot$ GCs. We model the expanding remnant made from the $10^{-4}M_\odot$ nova ejecta and show that it escapes long before it has cooled. We discuss the few positive ICM measurements and use a Monte-Carlo simulation of the accumulation and CNe recurrence times to reveal the possible variance in the ICM masses for the higher mass ($> 5\times10^5\ M_\odot$) GCs. We find that nova shells are effective at clearing the ICM in low-mass GCs ($\le 10^5\ M_\odot$), whereas higher-mass clusters may experience a quiescent time between novae long enough to prevent the next nova shell from escaping. The nova clearing mechanism will  also operate in ultra-faint Milky Way satellites, where many upper limits on gas masses are available. 
\end{abstract}


\section{Introduction}

The stars in a galactic globular cluster (GC) lose mass during their evolution, injecting material into the GC that accumulates over time to make an intra-cluster medium (ICM) of mass $M_{\rm ICM}$. The most robust mechanism for clearing this matter is ram pressure stripping during the GC's passage through the galactic disk every $\sim 10^8-10^9$ yr, implying $M_{\rm ICM}\sim 100-1000  M_\odot$ \citep{Tayler75}. However, searches for the ICM have yielded upper limits \citep{Birkinshaw83, Roberts88} or detections at levels more than 10 times smaller than implied by this mechanism \citep{Smith76, Freire, vanLoon}. Recent searches for dust in GCs confirm this paucity \citep{Lynch90, Knapp95, Origlia96, Penny97, Hopwood, Akari}. Spitzer observations place upper bounds on dust masses  $10-100 \times $ below that expected \citep{Barmby09}, and, where detections are made \citep{Boyer06, Boyer08}, the values are much lower than expected. 

It's unclear if the diversity in GC escape velocities and the possibility for fast stellar 
winds \citep{Dupree09} can explain all of these observations by simply having the material directly leave the cluster. A more robust solution for all GCs would be a more frequent ICM clearing mechanism than disk passages. Though many mechanisms (e.g. pulsar winds \citep{Spergel91}, stellar collisions \citep{Umbreit08}, ram pressure stripping by galactic halo medium \citep{Priestley10}) are discussed, one appears to be robust (but often neglected):  ICM clearing from classical novae (CNe) explosions \citep{SD}. \citet{SD} showed that CNe had explosion energies, $E_{\rm nova}$,  and recurrence rates adequate to unbind the accumulated ICM, but the astrophysical uncertainties (at that time) made definitive statements about this mechanism difficult. We show here  that the improved knowledge of mass loss from the stellar population; coupled with better constraints on the CNe rates and energetics reveals that CNe will clear the ICM and do so with a frequency that explains the ICM observations.  

The mass loss rate from the old stellar population in a GC, $\dot M_{\rm ICM}$,  is now observationally well constrained. The main sequence turnoff  mass  is $\approx 0.78-0.88 M_\odot$ \citep{Dotter09, Thompson10}  in an old ($>10$ Gyr) GC, and white dwarfs (WDs) are born with an average mass of $\approx 0.53\pm 0.01 M_\odot$ \citep{Kalirai09} implying $\approx 0.3M_\odot$ of material placed into the ICM.\footnote{ We do not distinguish between the matter lost on the
way to the Horizontal Branch (see Dotter 2009 for a recent discussion) versus that lost on the asymptotic giant branch.} We estimate the WD birthrate from the calculated specific evolutionary
flux for a Salpeter initial mass function (IMF) of $1.7\times 10^{-11} \ L_\odot^{-1} \ {\rm yr}^{-1}$ (from \citet{Buzzoni06}, where a steeper of shallower IMF only changes this number by  $\pm 20\%$) giving $\dot M_{\rm ICM} \approx 5\times 10^{-12}\ M_\odot  \ {\rm yr}^{-1}\ L_\odot^{-1}$. 
Assuming that  each CNe clears the accumulated ICM,  the typical ICM mass is  $M_{\rm ICM}\approx 0.03\ M_\odot$ for our fiducial (see \S 2) CNe rate of $20\ \rm{yr}^{-1}$ per $10^{11}\ L_\odot$. The  ICM binding energy in a GC of mass $M_{\rm GC}$ and  half-mass radius $R_h$ is 
\begin{equation}
E_{\rm b}\approx 10^{45} \ {\rm ergs}\left(M_{\rm GC}\over 5\times10^5\ M_\odot\right)\left(M_{\rm ICM}\over 0.03 M_\odot \right)\left(1 \ {\rm pc}\over R_h\right),
\end{equation} 
similar to the CNe energy release, $E_{\rm nova}$, as first highlighted  by \citet{SD}. 

These arguments motivate our re-consideration of the dynamics of CNe shell evolution in \S 3, where we show that the nova shells escape the cluster before suffering significant cooling.
In \S 4, we apply this model to all galactic GCs, and discuss the possibility that the current ICM knowledge can constrain the underlying population of binaries that cause CNe, the cataclysmic variables (CVs). We conclude in \S 5 and mention the impact that CNe will  have on the ICM in the dwarf spheroidal galaxies of the Milky Way's halo. 

\section{Classical Novae properties and rates}

CNe result from the unstable thermonuclear ignition of accumulated hydrogen on a WD in a mass-transferring binary \citep{Gehrz98}. The rapid nuclear burning drives  the ejection of most of the accumulated material, leading to outflows of $M_{\rm ej}\approx 10^{-4}-10^{-5}\ M_\odot$ at velocities of $V_{\rm ej}\approx 1000\ {\rm km \ s^{-1}} $ (see \citet{Starrfield72}; \citet{Prialnik86}). The typical kinetic energy is best inferred from observations late in the CNe
evolution (typically $> 10 $ years later \citep{Cohen83, Cohen85, Downes00}) where $V_{\rm ej}\approx 700-1200\ { \rm km \ s^{-1}}$ is measured for the shell, motivating a fiducial CNe explosion energy of
\begin{equation}
E_{\rm nova}={M_{\rm ej} V_{\rm ej}^2 \over 2}=10^{45} \ {\rm erg}\left(M_{\rm
ej}\over 10^{-4}M_\odot\right)\left(V_{\rm ej}\over 1000 \ {\rm km
\ s^{-1}}\right)^{2},
\end{equation} 
which we use throughout this paper. The calculated accumulated masses depend on the mass accretion rate, WD core temperature and accreted metallicity (\citet{Prialnik95}; \citet{TB}; \citet{Yaron05}) and, for a 'typical' case yield $M_{\rm ej}\approx 3(10)\times 10^{-5}\ M_\odot$ for
$M_{\rm WD}=1.0(0.6)\ M_\odot$, consistent with the observed range \citep{Gehrz98}. This variation in CNe ejecta masses motivates our later consideration of the impact of a range of $E_{\rm nova}$.  

The CNe rate in GCs is not known, as only two CNe have been observed in galactic GCs (see historical discussions in \citet{Shara04,Shafter07}), and three in extragalactic GCs (one of M87 \citep{Shara04}, and two of M31 \citep{Shafter07,Henze09,Henze10}). \citet{Shafter07} discussed this challenge, suggesting that the CNe rate per unit stellar mass in GCs may only be 
slightly higher than in the field, but is likely not much lower. Another way to make an estimate is to take 
advantage of the explicit connection between CNe and the observed CV population. 
 \citet{TB} showed that the majority of galactic CNe arose from CVs with $P_{\rm orb}<8 $ hours 
 undergoing the conventional mass transfer scenario of disrupted magnetic braking (see \citet{Howell01}). The measured 
CNe in distant galaxies,   $20\ \rm{yr}^{-1}$ per $10^{11}\ L_\odot$ \citep{Guth10} then implied a CV population per unit mass 
comparable to that seen in our local galactic disk. 

The combination of HST and Chandra observations of GCs are revealing their CV populations \citep{Edmonds03, Dieball05, Dieball07,Bassa08,Dieball10, Huang10}, allowing for much better raw statistics on their incidence. 
However, these CV populations  appear to be a combination of those from primordial binaries and dynamical interactions \citep{Pooley06, DiStefano}, and may, in the end, depend on the GC central stellar density, $\rho_0$.  \citet{Haggard09} discuss the situation in $\omega$ Cen, where they conclude a lower CV incidence per unit mass than in the field. Given this current diversity of answers at present, we simply 
adopt the field CNe rate of $20\ \rm{yr}^{-1}$ per $10^{11}\ L_\odot$ \citep{Guth10} for our work.\footnote{We can be confident that a GC will contain at least one CV since mass-transferring CVs have a birthrate of one every $400$ years per $10^{11}\ M_\odot$ \citep{TB}. If we assume an active lifetime of $3$ Gyrs, then one active CV requires $M_{\rm GC} > 10^4\ M_\odot$.\label{cvrare}} 

\section{Remnant Evolution and Escape}

We now show that the kinetic energy injected by a 
CNe, $E_{\rm nova}$, is adequate to unbind the ICM accumulated since the prior explosion.
Although local ICM structure will influence the shell evolution, our aim here is to show the ease of unbinding the ICM. We define a characteristic (and uniform) ICM number density $n_0\equiv 3M_{\rm ICM}/m_p4\pi R_h^3\simeq 0.1\ \rm{cm}^{-3}$, an overestimate since some ICM mass will be beyond $R_h$. We treat the evolution using a three-stage model \citep{Spitzer}, starting with the  brief free-expansion phase that lasts the time 
\begin{equation}
t_I = 120 \ {\rm yr} \left(\frac{M_{\rm{ej}}}{10^{-4}M_\odot}\right)^{5/6} \left(\frac{E_{\rm nova}}{10^{45}\ \rm{ergs}}\right)^{-1/2} \left(\frac{n_0}{\rm{cm^{-3}}}\right)^{-1/3},
\end{equation}
needed to sweep up the $M_{\rm ej}$ of ICM, creating a reverse shock. At this point the evolution transitions into the Sedov-Taylor phase and ends when the radiative cooling of the post shock material becomes significant. 

\begin{figure}[h!]
	\begin{center}
		\includegraphics[scale=0.7]{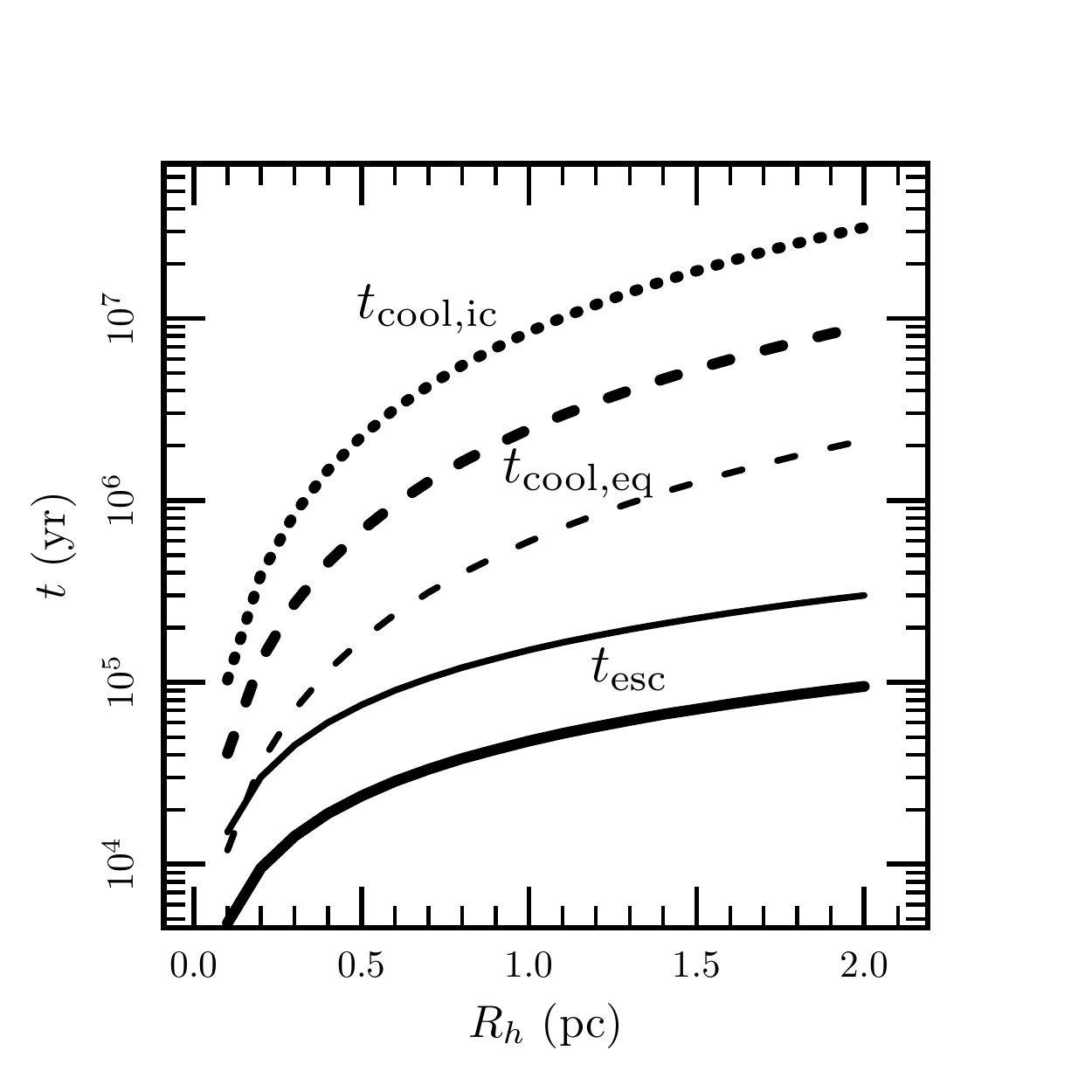}
		\caption{Timescales as a function of half-mass radius for a GC of mass $5\times 10^5 M_\odot$. The top three lines are various cooling times. The topmost dotted line is the non-equilibrium cooling time from isochoric cooling rates. The next two dashed lines are the equilibrium cooling times. The heavy line is corresponds to our fiducial $M_{\rm ICM}=0.03\ M_\odot$ and the lighter line to a $10{\rm x}$ higher $M_{\rm ICM}$ for comparison. The lowest two lines are the escape times for these same ICM masses, again with the fiducial in bold.
		\label{times}}
	\end{center}
\end{figure}

The simple case of a strong explosion in a homogeneous medium is given by the Sedov-Taylor solution \citep{Zel}. The radius of the shock front, $R_s$, velocity of the shock front, $V_s$, and temperature immediately behind the shock, $T_s$ are respectively, 
\begin{equation}
R_s(t) \approx 2.3\ {\rm pc}\left(\frac{E_{\rm nova}}{10^{45}\ \rm{ergs}}\right)^{1/5} \left(\frac{n_0}{\rm{cm^{-3}}}\right)^{-1/5} \left(\frac{t}{10^4 \ \rm{yr}}\right)^{2/5},
\end{equation}
\begin{equation}
V_s(t) \approx  92 \mbox{ km/s}\left(E_{\rm nova}\over 10^{45} {\rm ergs}\right)^{1/5}\left( n_0 \over {\rm cm}^{-3} \right)^{-1/5}\left(t \over 10^4 \ {\rm yr}\right)^{-3/5}, 
\end{equation}
\begin{equation}
T_s(t) = 1.1\times 10^5\mbox{ K}\left(E_{\rm nova}\over 10^{45} {\rm ergs}\right)^{2/5}\left( n_0 \over {\rm cm}^{-3}\right)^{-2/5}\left(t \over 10^4 \ {\rm yr} \right)^{-6/5},
\end{equation}
where $t$ is the time since explosion. These numbers show that  the shell will be moving at a speed in excess of the  GC escape velocity, $V_{\rm esc}\approx 30 \ {\rm km \ s^{-1}}$, even $10^4$ years after the explosion, consistent with our  energetic arguments.  The time it takes the shell to escape the cluster, $t_{\rm esc}$, is simply the time for the shell to reach $R_h$ 
with $V_s>V_{\rm esc}$. These are shown by the solid lines in Figure \ref{times}, where we have used the Plummer model \citep{BT} to estimate $V_{\rm esc}$. 

The only remaining question is whether the remnant radiates and cools prior to escape. To answer that, we 
calculate the cooling of the remnant using collisional ionization equilibrium cooling rates for a range of metallicities from \citet{GS}. The mass of the shell is heavily concentrated near the shock front, therefore we approximate the remnant as being spatially isothermal at the temperature just behind the shock, $T_s(t)$. From \citet{Cox}, the energy loss rate is 
\begin{equation}
\der{E}{t}(T,n)=-\int_0^{R_s}{ \Lam(T) n^2(R) 4\pi R^2 dR},
\end{equation}
where $\Lam$ is the cooling function (in erg cm$^3$ s$^{-1}$), so $\Lam n^2$ is the energy loss rate per unit volume. We adopt \citet{Cox}'s compaction parameter
\begin{equation}
\lam=\left(\frac{4}{3}\pi R_s^3\right)^{-1}\int_0^{R_s}{\left(\frac{n(R)}{n_0}\right)^24\pi R^2 dR},
\end{equation}
which has the value $\lam \simeq 2.3$ for a strong shock. This yields
\begin{equation}
\der{E}{t}=-\Lam n_0^2 \left(\frac{4}{3}\pi R_s^3(t)\right)\lam,
\end{equation}
an equation that gives the energy as a function of time, defining $t_{\rm cool}$, the time it takes for half the energy to be lost. We also compute the longer non-equilibrium cooling times for the case where the gas cools on a timescale shorter than the recombination times of the various ions,  causing a relative over-ionization as compared to the equilibrium case. We therefore distinguish cooling times that are calculated from the equilibrium cooling efficiencies, $t_{\rm cool,eq}$, and two types of non-equilibrium cooling efficiencies - isobaric, $t_{\rm cool,ib}$, and isochoric, $t_{cool,ic}$. The upper curves in Figure \ref{times} show the different cooling times for a range of $R_h$ for $M_{\rm GC}=5\times 10^5 M_\odot$ (and ICM particle densities ranging from $n_0=0.06-4\ \rm{cm}^{-3}$ implied by a fixed $M_{\rm ICM}=0.03M_\odot$). The isobaric and isochoric cooling times are similar so we only plot $t_{\rm cool,ic}$. All of these times are much longer than the escape time, making it clear that the shell will escape in the Sedov phase and radiative cooling will be an important energy loss mechanism only after the shell leaves the GC.

\section{Application to  Galactic Clusters}

\begin{figure}[h!]
	\begin{center}
		\includegraphics[scale=0.7]{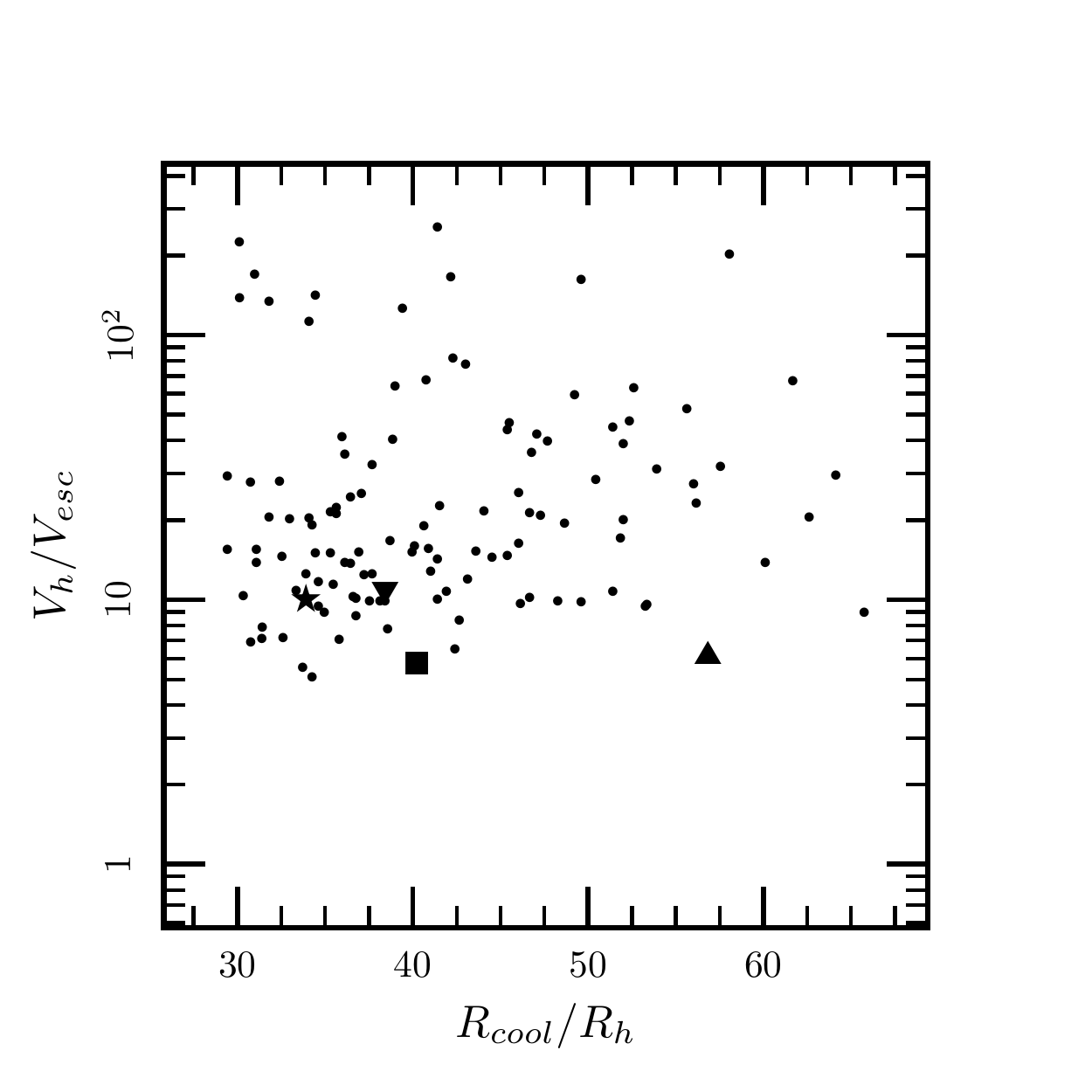}
			
		\caption{Outcomes for CNe driven remnants in Milky Way GCs. 
				The x-axis is the shell's radius at the cooling time, $R_{\rm cool}$, divided by the GCs half-mass radius, $R_h$.  The y-axis is the ratio of velocity of the shell at $R_h$ to the GC's escape velocity, $V_{\rm esc}$. The shapes indicate GCs with ICM observations: 47 Tuc is the up triangle; M15 the square; and M2 the down triangle. The starred cluster is M80, a Milky Way GC with an observed CNe (T Sco in 1860).   \label{vvsr}}
	\end{center}
\end{figure}

We now apply our model of ICM accumulation to the Milky Way GCs \citep{Harris96}. 
We assume each GC has accumulated $M_{\rm ICM}\approx 0.03\ M_\odot$ from stellar mass loss since the last clearing event, and 
that the characteristic particle density, $n_0$, is uniform. For each GC, we compute the equilibrium cooling time, the shell radius at the cooling time, $R_{\rm cool}$, and the shell's velocity, $V_h$, when it was at the half-mass radius (i.e. $V_h=V_s$ when $R_s=R_h$). We plot these values  in Figure \ref{vvsr}, showing that $V_h>V_{\rm esc}$ in all cases, and that cooling occurs when $R_s\gg R_h$. Escape is the outcome in this simple formulation. 

Since we don't expect the CNe to be strictly periodic, we performed a Monte Carlo simulation for $10^8$ years allowing for variation in CNe recurrence times and drawing the nova energies from a log-uniform distribution 
ranging from $2.5-10\times10^{44}$ ergs. For each GC, we calculate the expected number of CVs (assuming $1$ CV per $10^{4}\ M_\odot$, see footnote \ref{cvrare}), and generate a nova explosion history assuming an average CNe recurrence time in the range $(3.3-6.7)\times 10^5$ yrs, centered on 
the recurrence time calculated from the CNe rate, $5\times 10^5$ yrs. 

We calculate how large an effect these variances will have by calculating how often a GC is in a state where $M_{\rm ICM} > 0.03\ M_\odot$. During the $10^8$ years, there is a $12\%$ chance that the GC will have twice the fiducial $M_{\rm ICM}$, a $4\%$ chance for $3\times$, and a $0.4\%$ chance to have $5\times$ the fiducial ICM built up - a prediction nearly independent of GC parameters. Shell escape can become difficult for some galactic GCs if $M_{\rm ICM}$ is increased by a factor of $10$. Figure \ref{vvsr} looks qualitatively the same in that case, but the line representing $V_h = V_{\rm esc}$ would now cross between the triangles on the y-axis and the x-axis range would shrink to $6-14$. 

The maximum quiescent time was $\sim 10\times$ the average time between novae, allowing for a very large buildup of $M_{\rm ICM}$, potentially to values where $E_{\rm b} > E_{\rm nova}$,  forcing us to consider the possibility that some GCs may, due to the randomness of the process, 
experience a ``runaway" accumulation. We determine the probability for this outcome by 
computing the average time it takes for this condition to be reached and assuming that the GC has uniform probability of being at any point in its $10^8$ yr orbital period (longer orbital periods will increase the chance of seeing a runaway). For $\rho_0 = 10^3\ L_\odot/{\rm pc}^3$ and $R_h= 2.0 {\rm  pc}$, we find that  GCs with $M_{\rm GC}=(1.5-5)\times 10^5\ M_\odot$ have an even chance of experiencing a runaway, where the exact value depends on the nova energy distribution.  This probability increases very quickly with mass, so more massive clusters are much more likely to undergo a runaway while less massive GCs will almost never have one. This probability is also sensitive to $\rho_0$, with more collapsed clusters much more likely to undergo runaways. Although we refer to it as a ``runaway", it may not be unmitigated mass accumulation, as the asymmetries of the ICM will allow the CN to unbind some material. However, the stochastic nature of this process inhibits our ability to make a secure prediction for a specific GC. 
 
There are also some instances (e.g. M15) where the shell escape time, $t_{\rm esc}$, is so long that it does not escape before another CN occurs. The outcomes from these colliding remnants 
are more complex, either leading to a net heating of the ICM that creates a wind like solution \citep{SD} or more intense cooling that keeps the material bound. In either case, 
novae clearing is most effective in the smaller, less dense clusters, where the shell safely 
escapes before another nova goes off and the GC is in no danger of a runaway scenario.

There have been few positive ICM measurements in GCs. \citet{Barmby09} derive $M_{\rm ICM}$ upper limits of $0.04-1.4\ M_\odot$ from dust mass upper limits. These upper limits are (with the exception of M2) consistent with observed $M_{\rm ICM}$ values from \citet{Freire}, \citet{vanLoon}, and \citet{Grindlay77}. We collect the four reported $M_{\rm ICM}$ measurements in Table \ref{GC Data}, along with other useful GC parameters and timescales we derive. The ranges in $t_{\rm esc}$ and $t_{\rm cool}$ come from taking our fiducial $M_{\rm ICM}=0.03\ M_\odot$ in the clusters for one bound (the upper bound on $t_{\rm cool}$ and the lower bound on $t_{\rm esc}$) and the measured $M_{\rm ICM}$ for the opposite bounds. Selection effects will bias likely detections towards GCs with higher $M_{\rm ICM}$ values.  According to our calculations, the high central densities and masses of these clusters place them in the runaway regime. 

We have assumed that the mass lost from stellar evolution is a smooth process at the rate of $\dot M_{\rm ICM} \approx 5\times 10^{-12}\ M_\odot  \ {\rm yr}^{-1}\ L_\odot^{-1}$. However, stochastic effects in the mass loss should be examined as well.
We start by considering the most extreme case, where the mass is lost in a sudden event (e.g. He core flash, thermal pulse on the AGB). Since the WD birth rate is one-tenth the CNe occurrence rate, such an event would occur on an average timescale of $10t_{\rm rec}$. Though certainly the case that the CNe immediately following this event would not be adequate to unbind it (both because of energetic arguments and relative filling fractions), such a stellar ejection ``chunk" would be shocked, on average, by  $\approx 10$ CNe. Whether this proves sufficient to unbind it depends on the location and time to cool between shocks; which is beyond the scope of our work here. 

An additional possibility is wind-driven mass loss that is so strongly dependent on the stellar luminosity (e.g. $\dot M\propto L^\alpha$), that it makes $\dot M_{\rm ICM}$ time dependent on the $t_{\rm rec}$ timescale. As an example,  consider mass-loss near the tip ($L_{\rm TRGB}$) of the RGB. It takes $3\times 10^6$ yrs to evolve from $L=L_{\rm TRGB}/2$ to $L=L_{\rm TRGB}$ \citep{Salaris97}, implying $\approx 30$ stars in these final stages for a $5\times 10^5\ M_\odot$ cluster. Let's consider the amount of mass one of these stars could lose during the final $t_{\rm rec}$ of its evolution to the expected total mass loss along the RGB, $\sim 0.2 M_\odot$ \citep{Origlia07}. For typical values of $\alpha\sim 1-3$ \citep{McDonald09,Meszaros09}, this ratio is less than $10^{-3}$. To get a substantial mass loss in the last $t_{\rm rec}$ requires $\alpha > 100$, basically becoming equivalent to the sudden event model already discussed.  So, we do not expect any of the current wind models or observations to create a time dependent $\dot M_{\rm ICM}$  on the CNe recurrence timescale. 

\section{Conclusions and other applications}

We have reconsidered the \citet{SD} model of using CNe explosions to clear the accumulated ICM  in galactic GCs, allowing for a more secure analysis of its efficacy. Using estimated CNe rates and the  known stellar mass loss rates, we predict that novae clearing is highly effective 
in the less massive and less centrally condensed galactic GCs (those unlikely to undergo a runaway), leading to a typical ICM buildup of $\approx 0.03M_\odot$ that is consistent with observational upper limits. For those few GCs with positive ICM measurements, the $M_{\rm ICM}$ values are higher than our predictions, though still 1-2 orders of magnitude less than the 
$\sim 300\ M_\odot$ predicted from galactic plane crossing. We discuss in \S 4 whether this could simply be the result of the stochastic nature of CNe explosions and the large number of GCs observed, or from a systematic dependence on $M_{\rm GC}$ or $\rho_0$. In either case, it is clear
that CNe have a large deleterious impact on the ability for the GC to build up it's ICM. 

A very robust mechanism to clear the ICM are Type Ia supernovae. The simplest assumption \citep{Pfahl09} is to use the rate expected from the mass, $(5.3\pm 1.1)\times 10^{-14}\ {\rm yr^{-1}}\ M_\odot^{-1}$ \citep{Sullivan06}, yielding an average time between events of $\approx 2\times10^7\ {\rm yrs}$ in a $10^6\ M_\odot$ GC, shorter than a galactic plane crossing. 
If CNe are ineffective in the intervening time, then the ICM mass would reach $M_{\rm ICM} \approx 100\ M_\odot$,   but with a binding energy orders of magnitude less than the $10^{51}$ ergs of the SNe. Such outcomes may be more prevalent in massive GCs around elliptical galaxies, since there is no plane crossing to clear out accumulated ICM. When $M_{\rm ICM} \approx 100\ M_\odot$ the explosion would occur in a  
density of $n_0\approx 100\ {\rm cm^{-3}}$, much larger than expected in an elliptical galaxy, $\approx 10^{-2}\ {\rm cm^{-3}}$ \citep{Stewart84}. Application of our work in \S 2 finds that the velocity of such a remnant would be 
$V_s\approx 130 \ {\rm km \ s^{-1}}$ at a time of 2700 years, when it reaches $R_h= 1$ pc. The cooling time is orders of magnitude longer, and so this shell will easily escape, and be quite bright (due to the high density).  Less than a few $\%$ of
extragalactic GCs have shown emission lines consistent with these velocities \citep{Chomiuk08}, raising the question as to whether these could  be the first indicators of the expected Ia explosions in GCs. If such a remnant is 
detectable for $10^4$ years, then we would expect to see one such case amongst all the GCs in a large elliptical galaxy like  M87. 

\begin{center}
\begin{deluxetable}{lcccccccr}
\tabletypesize{\scriptsize}
\tablecaption{GCs with ICM detections \label{GC Data}}
\tablewidth{0pt}
\tablehead{
\colhead{Cluster} & \colhead{$M_{\rm GC}$} & \colhead{$R_{h}$} & \colhead{$[{\rm Fe/H}]$} &
\colhead{$M_{\rm ICM}$} & \colhead{$E_{\rm bind}$} & \colhead{$t_{\rm rec}$} &
\colhead{$t_{\rm esc}$} & \colhead{$t_{\rm cool}$} \\
\colhead{name} & \colhead{$(10^5\ M_\odot)$} & \colhead{$({\rm pc})$} & \colhead{(dex)} &
\colhead{$(M_\odot)$} & \colhead{$(10^{45}\ {\rm ergs})$} & \colhead{$(10^3\ {\rm yr})$} &
\colhead{$(10^5\ {\rm yr})$} & \colhead{$(10^8\ {\rm yr})$}
}
\startdata
47 Tuc  \tablenotemark{1} & $15$ & $2.8$ & $-0.8$ & $0.1$ & $0.5$ & $3$ & $1-2$ & $0.5-1$ \\
M2  \tablenotemark{2} & $8.8$ & $0.9$ & $-1.6$ & $3$ & $100$ & $6$ & $0.4-4$ & $0.07-1$ \\
M15  \tablenotemark{2} & $5$ & $6.4$ & $-2.3$ & $0.3$ & $40$ & $10$ & $3-10$ & $1-5$ \\
NGC 6624 \tablenotemark{3} & $3.3$ & $0.8$ & $-0.4$ & $0.02$ & $0.5$ & $15$ & $0.3-0.4$ & $2$
\enddata

%% Text for table notes should follow after the \enddata but before
%% the \end{deluxetable}. Make sure there is at least one \tablenotemark
%% in the table for each \tablenotetext.

\tablenotetext{1}{\citet{Freire}}
\tablenotetext{2}{\citet{vanLoon}}
\tablenotetext{3}{\citet{Grindlay77}}

\end{deluxetable}
\end{center}

Novae clearing can also be applied to the newly discovered ultra-faint (only $\approx 10^4M_\odot$ in stars, potentially so low that only a few CVs may be present, see footnote \ref{cvrare}) dwarf spheroidals \citep{Belokurov07,Simon07}, where ICM deficits have been reported \citep{Bailin07}.\footnote{The large HI mass (comparable to the stellar mass of $2\times 10^5M_\odot$) measured by \citet{Ryan08} in the distant dSpH Leo T \citep{Irwin07} exceeds that possible from stellar wind accumulation over 10 Gyrs, and is likely from continued infall of material. Such an inflow may also be the source of the 10$\%$ of stellar mass in young ($\approx $ Gyr) stars \citep{DeJong08}.} Just as in GCs, these dwarf spheroidals (dSph)  have a mass-losing stellar population. However, the lower escape velocities ($2-5 \ {\rm km/s}$ from the $\approx 10^6M_\odot$ of dark matter in a half-mass radius of 100 pc) may allow the stellar winds to directly leave the dSph. In cases where the mass accumulates, the ICM binding energy is $< 10^{45} {\rm ergs}$ for $M_{\rm ICM} = 0.03\ M_\odot$, so that a single CNe can be effective. Though no challenge energetically, the 100 pc dimension means that the escape timescale is so long ($\ge 10^5 {\rm yrs}$) that much of the ICM would still be present when the next nova occurs. These systems are therefore much closer to the original limit discussed by \citet{SD}, which predicts a wind leaving the system at the expected $\dot M_{\rm ICM}$, but with a kinetic energy flux given by $E_{\rm nova}$ and the CNe recurrence time. Such a wind would have a terminal velocity of $\approx 60 \ {\rm km \ s^{-1}}$ (independent of the dSph mass), and a density at 100 pc of $\sim 10^{-7} {\rm cm^{-3}}$ for a dSph with $10^4M_\odot$ of old stars. 

We thank Matt van Adelsberg, David Chernoff, Aaron Dotter, Craig Heinke, David Kaplan, Michael Shara, Ken Shen, and Jay Strader for discussions. We also thank the anonymous referee for their useful comments.
This work was supported by the National Science Foundation under grants PHY 05-51164 and AST 07-07633.

\newpage

%\bibliographystyle{apj}
%\bibliography{Paper}

\end{document}